# Did Gravitational Waves Affect the Evolution of the Universe?

**David Garrison**
*Physics Department, University of Houston Clear Lake, 2700 Bay Area Blvd., Houston, Texas 77058*

Email: garrison@uhcl.edu

**Abstract.** As Cosmologists struggle to understand the evolution of our universe, an often over-looked element is the affect of gravitational radiation on the primordial plasma field. This appears to be due to the inaccurate approximation of the primordial plasma field as a simple hydrodynamic fluid, therefore neglecting the full dynamics of the magneto-fluid. In this paper, we show how gravitational waves may have had a significant impact on the evolution of our universe. While more work is needed, the author hopes to eventually develop a large-scale simulation of structure formation in the early universe, which takes the interactions of scalar perturbations, gravitational waves, magnetic fields and a dynamic plasma field into account.

PACS: 47.75.+f, 95.30.Sf, 98.80.-k

## 1. Introduction

Our knowledge of how the universe evolved comes primarily from observations of large structures such as stars, galaxies, clusters and super-clusters of galaxies as well as from observations of the Cosmic Microwave Background (CMB) Radiation. Based on these observations, the Standard Model of Cosmology was developed during the mid to late twentieth century. Some elements of this model include the existence of primordial metric perturbations and an early universe filled with a nearly homogenous and isotropic plasma field [16]. The Perturbed Friedmann-Robertson-Walker (FRW) metric, which describes the space-time curvature of the early universe, takes the following form,

$$ds^2 = -(1+2\varphi)dt^2 + \omega_i dx^i dt + a^2(t)\left[((1+2\psi)\delta_{ij} + h_{ij})dx^i dx^j\right]. \qquad (1.1)$$

Here $a(t)$ is the scale factor and $\varphi$, $\psi$, $\omega_i$, and $h_{ij}$ are the scalar, vector and tensor perturbation terms. Many cosmological models relate density fluctuations and variations in the CMB to perturbations in the FRW metric at the time of recombination. These perturbations start off small and grow as a power-law with time as the competing forces of universal expansion and gravitational attraction affect their growth [16]. Work by Kodama and Sasaki [14], Sachs and Wolfe [23] and Mukhanov, Feldman and Brandenberger [16] all showed how metric perturbations could cause density perturbations in a hydrodynamic fluid. Unfortunately, they (and almost every author who followed them) all approximated the plasma field as a hydrodynamic fluid instead of as a Magnetohydrodynamic (MHD) fluid. This approximation may have first appeared in order to simplify the problem to fit within the computational capacity of the day or because little was known about the dynamics of the magnetofluid and primordial magnetic fields. For whatever reason, interesting physics may have been lost because of this approximation. As a result, it became a common belief in cosmology that only scalar perturbations could significantly affect the fluid and form density fluctuations. This approximation quickly became a pillar of mainstream cosmological theory and therefore its validity was rarely questioned.

In section 2, we review a theory about how tensor perturbations (or gravitational waves) may have been amplified in the early expanding universe, making them strong enough to significantly interact with the primordial plasma field. Note that scalar perturbations should undergo a similar amplification process. In section 3, we review a theory that attempts to explain the origin of matter-antimatter asymmetry as a result of gravitational waves. This results in birefringent gravitational waves, which could play an interesting role in the dynamics of the magnetofluid. In section 4, we introduce two theories of cosmic structure formation. The first and most accepted is the Gravitational Instability Theory, which directly relates scalar metric perturbations to density fluctuations and structure formation. The second theory, which last appeared about 30 years ago, is the Cosmological Turbulence Theory that tried to explain structure formation as the result of turbulence in the early universe. In section 5, we introduce Magnetohydrodynamic turbulence and the surprising result that a magnetofluid experiencing turbulence tends to form coherent structures. This section will further explain why approximating a primordial magnetofluid as a hydrodynamic fluid doesn't work well. In section 6, we discuss the current state of knowledge of primordial magnetic fields. The existence of these fields affects the dynamics of the primordial plasma field and their interaction with gravitational waves. In section 7, we present work on how gravitational waves may have affected the formation of cosmic structures through their interactions with the primordial plasma field as well as preliminary results supporting this theory. In section 8, we conclude.

## 2. Amplification of Primordial Gravitational Waves

According to work by Grishchuk [10], the spectrum of gravitational waves from the early universe can be calculated by treating the system of quantum fluctuations in an expanding universe like a variable-frequency harmonic oscillator. The first step in doing this is to re-write the wave equation for a gravitational wave in a form similar to the Schrödinger equation

$$\mu'' + \mu\left[n^2 - \frac{a''}{a}\right] = 0. \quad (2.1)$$

Here n is the dimensionless wave number, $\mu$ is related to the amplitude of the gravitational wave by $h(\eta,\bar{x}) = [\mu(\eta)/a(\eta)]e^{i\bar{n}\cdot\bar{x}}$ and $U(\eta) = a''/a$ acts as an effective potential caused by the expanding space-time. $\eta$ is conformal time defined by the equation $\eta = -1/Ha$, which ranges from negative infinity to nearly zero. If $n^2$ is much larger than the magnitude of the effective potential, the wave's amplitude drops adiabatically like the inverse of the scale factor. If $n^2$ is close to the effective potential the amplitude is instead amplified beyond the adiabatic law. Grishchuk calls this process "super-adiabatic amplification".

Super-adiabatic amplification relies on only general relativity and quantum mechanics to be valid in their current forms. It does not rely on any new physics or modifications to these theories. According to Grishchuk [10], relic gravitational waves start as zero-point quantum fluctuations inside a cosmological system that behaves similarly to a classical pendulum. The variable effective frequency of perturbations in the cosmological system is similar to the variable length (and therefore the natural frequency) of a pendulum. In cosmology, the effective frequency $n^2 - a''/a$ of the cosmic oscillation changes because of the variable external gravitational field represented by the scale factor $a(\eta)$. This results in the parametric amplification of not only scalar but also tensor metric perturbations.

The unique thing about this theory is that it does not rely on the assumption of a preset tensor/scalar ratio as is commonly used in cosmological theory. Grishchuk's theory therefore provides predictions for the spectrum of relic gravitational waves from the early universe. This theory also predicts that these relic gravitational waves should be observable with future gravitational wave interferometers such as LISA. So far the predictions of this theory are consistent with cosmic microwave background (CMB) observations [10]. According to Grishchuk the spectrum of gravitational wave perturbations is given by

$$h(n,\eta) = 8\sqrt{\pi}l_{pl} |1 + \beta(\eta)|^{-(1+\beta(\eta))} n^{2+\beta(\eta)} / l_0. \qquad (2.2)$$

Where $l_{pl}$ is the Planck length, $\beta(\eta)$ is a parameter related to slow-roll inflation and $l_0$ is a coefficient for power-law inflation. $\beta(\eta)$ and $l_0$ are related to the scale factor by,

$$a(\eta) = l_o |\eta|^{1+\beta} \quad -\infty \leq \eta < 0; -2 \leq \beta \leq -1 \qquad (2.3)$$

## 3. Birefringent Gravitational Waves

Work by Alexander, et all [1] uses gravitational waves to explain the asymmetry between matter and antimatter (This asymmetry could also be a nonlinear phenomenon resulting from the dynamics of the magnetofluid [25]). The WMAP experiment has confirmed that there is an excess of baryon density to photon density

$$\frac{n_B}{n_\gamma} = (6.5 \pm 0.4) \times 10^{-10}, \qquad (3.1)$$

where $n_B$ is the difference between baryon and anti-baryon number density and $n_\gamma$ is the number density of photons. This observation confirms what we know about the relative abundances of matter and anti-matter in the universe but leads us to the question, why is the universe made mostly out of matter? According to our accepted view of fundamental physics, matter and anti-matter are made in pairs from the decay of high-energy photons. An excess of matter over anti-matter should require some sort of symmetry-breaking mechanism.

Alexander's theory, called Gravi-leptogenesis, utilizes the three Sakharov Conditions to explain this symmetry-breaking: 1) baryon number violating vertices are present 2) CP is violated and 3) both violations are relevant when the universe is not in thermal equilibrium. The Standard Model of particle physics loop suppresses baryon number violating interactions and does not provide enough CP violations while out of equilibrium to satisfy the Sakharov Conditions. However, an asymmetry in baryons may have grown from an asymmetry in leptons. Unfortunately, the Standard Model with massless neutrinos does not satisfy the Sakharov Conditions for leptons any better than for baryons so physics beyond the Standard Model is needed.

Gravi-leptogenesis uses gravitational fluctuations created in the early universe to satisfy the Sakharov Conditions. For example, the Lepton number violation is explained by a triangle anomaly, which relates fermion current to a gravitational anomaly in the Standard Model.

$$\partial_\mu J_l^\mu = \frac{N}{16\pi^2} R\hat{R}$$

$$J_l^\mu = \sum_{i=L,R} \bar{l}_i \gamma^\mu l_i + \bar{\nu}_i \gamma^\mu \nu_i, \qquad R\hat{R} = \frac{1}{2}\varepsilon^{\alpha\beta\gamma\delta} R_{\alpha\beta\rho\sigma} R_{\gamma\delta}{}^{\rho\sigma}$$

(3.2)

Here $N = N_L - N_R$, which is 3 in the Standard Model. The anomaly is a result of a difference in the number of left- and right-handed leptons, leading to a nonzero value for lepton current. On the right-hand side of the equation are the curvature terms. These are a manifestation of the CP violation where lepton generation only occurs when $\langle R\hat{R} \rangle$ is non-vanishing. This happens during inflation when the inflation field has an odd CP component giving a non-zero asymmetry parameter.

By adding the interaction for coupling the axion to gravity to the action that typically classifies the generation of gravitational waves, Alexander determined the equations of motion for these waves. If the action is written in terms helicity where

$$h_L = \frac{1}{\sqrt{2}}(h_+ - ih_\times), \quad h_R = \frac{1}{\sqrt{2}}(h_+ + ih_\times),$$

(3.3)

represent left and right-handed gravitational waves respectively, it can be shown that $R\hat{R}$ vanishes if $h_L$ and $h_R$ have the same dispersion relationship. Therefore $R\hat{R}$ requires that we have "cosmological birefringence" during inflation. The equations of motion for $h_L$ and $h_R$ along the z-axis then become

$$(\frac{\partial^2}{\partial t^2} + 3H\frac{\partial}{\partial t} - \frac{1}{a^2}\frac{\partial^2}{\partial z^2})h_L = -2i\frac{\theta}{a}\frac{\partial^2}{\partial t \partial z}h_L, \quad (\frac{\partial^2}{\partial t^2} + 3H\frac{\partial}{\partial t} - \frac{1}{a^2}\frac{\partial^2}{\partial z^2})h_R = +2i\frac{\theta}{a}\frac{\partial^2}{\partial t \partial z}h_R. \quad (3.4)$$

$\theta$ is determined by the interaction coupling the axion to gravity. Setting $\theta$ to zero yields the standard equations of motion for primordial gravitational waves. This result may be important to cosmological structure formation because any change in the polarization or spectrum of the primordial gravitational waves could have an impact on the dynamics of their interaction with the plasma field. To see how this affects the spectrum predicted by Grishchuk, see the paper by Garrison and de la Torre [8].

**4. Two Theories of Cosmic Structure Formation**

It is well accepted that linear tensor perturbations of the metric do not couple to the energy and pressure of a hydrodynamic fluid [16]. Because of this they cannot contribute to the gravitational instability needed to form cosmic structures. In addition, vector perturbations decay kinematically in an expanding universe making them ineffective for structure formation. This led cosmologists to see scalar perturbations as the most likely vehicle for growing cosmic structures. The Gravitational Instability Theory [13] of cosmic structure formation was a direct result of this logic.

According to the Gravitational Instability Theory, small deviations in the cosmological model can be evolved using linear perturbation theory. If there are no pressure gradients, small regions with a higher density than the surrounding fluid will expand more slowly than the rest of the universe. As a result the density contrast will grow with time. In the absence of an expanding space-time, these regions would collapse and the densities would increase exponentially. Because of the space-time expansion, the change in density should follow a power-law until pressure forces balance the perturbed gravitational field.

The Jeans length and the damping scale limit the scales at which structures are formed by gravitational instabilities at the high and low ends respectively. The Jeans length scale is the minimum length where pressure gradients can balance the gravitational forces. In a uniform fluid sphere, where in the absence of pressure the free-fall time for gravitational collapse is $t_f$, the Jeans length is $c_s t_f$. Here $c_s$ is the speed of sound in the fluid. A volume of fluid is stabilized from collapse if a sound wave can cross the volume on a time shorter than the collapse timescale.

The damping scale is relevant because it is not possible to propagate perturbations with an arbitrarily small wavelength through physical space. The wavelength is limited by the viscosity and thermal conductivity of the cosmic fluid. Both processes remove energy from high frequency sound waves. The damping scale is therefore the shortest wavelength whose damping time scale from dissipative processes is longer than the cosmic expansion time.

One weakness of the Gravitational Instability Theory is its explanation of the origin of galactic rotation. According Hoyle [10], galaxies acquire spin as the result of tidal stresses from external objects. Another idea from Doroshkevich [5] and Zel'dovich and Sunyaev [27] is that galaxies formed when a protocluster of galaxies collapsed. Taken together, these ideas imply that isolated galaxies, if they exist, should have less angular momentum than cluster galaxies. This would also imply that clusters of galaxies should only have angular momentum if they are in the vicinity of another galactic cluster.

The Cosmic Turbulence Theory was originally introduced during the 1940's shortly after the development of the theory of hydrodynamic turbulence [13]. It appeared to be inspired by the philosopher Epicurus [13] who believed that the primordial state of the universe was chaos and order later developed. Although the theory showed promise and no refute to the turbulence theory was published, it seemed to disappear in the mid 1950's. One outstanding issue was that the origin of the turbulence had yet to be explained. Another issue was whether or not the turbulence could survive long enough to create structures when dissipative forces where present. Ozernoi later reintroduced the theory in 1968 [18,19,20,21,22].

Ozernoi's theory differs from the original in two important ways. First, as the universe expands the large scale turbulent motions are "frozen out" therefore decreasing the turbulence scale. This freezing out process stops the turbulent eddies from decaying. Second, the speed of sound in the cosmic medium starts off as a significant fraction of the speed of light but then falls to only about *3 km/s* after recombination [13]. This causes random motions to become hypersonic around the time of recombination resulting in isothermal shocks and large density fluctuations. This short period of time will cause enormous compressions and dense lumps of material to form.

One advantage to turbulence is that it allows for a "turbulent energy cascade" where energy is transferred from large-scale dynamical motions to smaller scales as a result of collisions between streams of fluid. This allows the system to self-regulate by limiting the size of large structures while generating smaller structures. Because of viscosity a large eddy will only persist for about one rotation before its energy is completely transferred to smaller fluid motions and heat. If however the universe expansion time scale is shorter than the eddy turnover time scale, the eddy is "frozen in". A frozen eddy is not expected to lose energy to the turbulent energy cascade.

The problem with this theory is that the "supersonic" regime after recombination would result in dense bound lumps of materials that do not resemble galaxies. The advantage of the theory is that it can predict the masses and rotations of galaxies but this requires that the protogalaxies correspond to eddies with time scales shorter than the cosmic expansion time scale. If the eddies turned slower than the cosmic expansion time scale, the theory would fail to explain the origin of galaxies.

Both the Gravitational Instability Theory and the Cosmic Turbulence Theory refer to dynamical processes which occur after *t = 1 second*. The flaw in both this theories is that they assume a hydrodynamic stress energy tensor,

$$T^{\mu\nu} = \rho_0 h u^\mu u^\nu + P g^{\mu\nu}. \quad (4.1)$$

Rather than a MHD stress energy tensor,

$$T^{\mu\nu} = (\rho_0 h + b^2) u^\mu u^\nu + (P + \frac{b^2}{2}) g^{\mu\nu} - b^\mu b^\nu. \quad (4.2)$$

Here $\rho_0$ is the rest-mass density, $u^\mu$ is the four-velocity of the fluid and $h$ is the specific enthalpy related to the specific internal energy $\varepsilon$ by $h = 1 + \varepsilon + P/\rho_0$. $P$ is pressure and $b^\mu$ is derived from the magnetic field as measured by a normal observer moving with a four-velocity $n^\mu$,

$$b^\mu = -\frac{(g^\mu_\nu + u^\mu u_\nu) B^\nu}{\sqrt{4\pi} n_\nu u^\nu}. \quad (4.3)$$

The author believes that the truth of cosmic structure formations exists somewhere between the Gravitational Instability and Cosmic Turbulence theories but also involves dynamics unique to the magnetofluid. In the next two sections we will discuss how the dynamics of MHD turbulence is different than that of hydrodynamic turbulence.

**5. Magnetohydrodynamic Turbulence**

Work by Shebalin [24,26], on Homogeneous MHD Turbulence best demonstrates how the dynamics of a magnetofluid can differ from that of a hydrodynamic fluid. Plasma can more accurately be modeled as a fluid made up of charged particles that are therefore affected by magnetic fields as well as particle-particle interactions. Because of this, the magnetic field becomes a dynamic variable in addition to density, pressure and the velocity of particles. For example, In MHD turbulence, an equipartition occurs and we expect kinetic and magnetic energy fluctuations to become roughly equal. Shebalin modeled the magnetofluid as a homogenous system where the same statistics are considered valid everywhere in the computational domain. He utilized periodic boundary conditions and spectral methods in order to study how the dynamics of different scales interacted without the addition of boundary errors. Much of his work focused on an Ideal MHD system, where the magnetic and fluid dissipation terms were excluded.

By varying the mean magnetic field *(B$_o$)* and angular velocity *($\Omega_o$)* of the system, Shebalin was able to define five different cases with different invariants. In such a system there could be as many as 3 invariants; energy *(E),* and the psuedovectors cross helicity *(H$_c$)* and magnetic helicity *(H$_m$)*. In addition, the invariant parallel helicity, *H$_P$*

$= H_C - \sigma H_M$ ($\sigma = \Omega_o/B_o$), can be formed from a linear combination of cross and magnetic helicity. In a hydrodynamic fluid the only invariant is energy.

Table 1. MHD Turbulence and Invariants

| Case | Mean Field | Angular Velocity | Invariants |
|------|------------|------------------|------------|
| I | 0 | 0 | $E, H_C, H_M$ |
| II | $\mathbf{B}_o \neq 0$ | 0 | $E, H_C$ |
| III | 0 | $\mathbf{\Omega}_o \neq 0$ | $E, H_M$ |
| IV | $\mathbf{B}_o \neq 0$ | $\mathbf{\Omega}_o = \sigma \mathbf{B}_o$ | $E, H_P$ |
| V | $\mathbf{B}_o \neq 0$ | $\mathbf{\Omega}_o \neq 0$ ($\mathbf{B}_o \times \mathbf{\Omega}_o \neq 0$) | $E$ |

For an incompressible fluid $u(k,t)$ is the Fourier coefficient of turbulent velocity and $b(k,t)$ is the Fourier coefficient of the turbulent magnetic field. The energy, cross helicity and magnetic helicity can be expressed in terms of these as:

$$E = \frac{1}{2N^3} \sum_k [|u(k)|^2 + |b(k)|^2]$$

$$H_c = \frac{1}{2N^3} \sum_k u(k) \cdot b^*(k) \quad (5.1)$$

$$H_m = \frac{1}{2N^3} \sum_k \frac{i}{k^2} k \cdot b(k) \times b^*(k)$$

Phase portraits resulting from computer simulations of the five runs show that coherent structures formed in many systems where the magnetofluid was experiencing turbulence [24,26]. Coherent structure is defined as a system where variables with the smallest eigenvalues have large mean values. The cause of this structure formation appears to be a hidden broken ergodicity in the canonical theory. Although the governing equations are symmetric under symmetry transformations, this symmetry must be broken because of the psuedovectors. A system that conserves helicity with a positive sign cannot conserve it with a negative sign and vice versa. This appears to be true for systems with real turbulence as well. In a hydrodynamic system, this would not be the case because helicity in not an invariant in a hydrodynamic system. Shebalin's results imply that if the plasma field in the early universe was turbulent the result should have been the formation of coherent structures. Did this happen? If so, what would these structures look like?

## 6. Primordial Magnetic Fields

The biggest difference between a hydrodynamic fluid and a magnetofluid is how they are affected by magnetic fields. It follows that if there were no magnetic fields in the early universe, the primordial plasma could be appropriately approximated by a hydrodynamic system. However, if magnetic fields were present in the early universe then a magnetohydrodynamic system is needed to understand the evolution of the early universe. In this section, we discuss what is known about primordial magnetic fields and how they may have grown into the large intergalactic fields that we observe today.

The only thing that we really know about magnetic fields in the early universe is that little is known about their existence or absence. There are no direct observations of primordial magnetic fields. Our only knowledge of magnetic fields in this epoch comes from theories that fail to agree on the strength of the primordial magnetic fields. There are currently several dozen theories about the origin of cosmic magnetic fields [3,9]. The main reason that we believe that primordial magnetic fields existed is because they may have been needed to seed the large magnetic fields observed today. These seed fields may have also affected structure formation. Most theories of cosmic magnetic field generation fall into one of three categories [3,4,9].

1) Magnetic fields generated by phase transitions
2) Electromagnetic Perturbations expanded by inflation
3) Turbulent MHD resulting in charge and current asymmetries

Most models calculate the magnitude of primordial magnetic fields by starting with the observed strength of galactic or intergalactic magnetic fields and calculating how this field should have been amplified or diffused by external effects such as the galactic dynamo and expansion of the universe [3,9]. A major problem is that there doesn't appear to be a universal agreement of how efficiently a galactic dynamo could have strengthened seed magnetic fields. Estimates of the strength of these seed fields can vary by tens of orders of magnitude. If there are no amplification mechanisms, the frozen-in condition of magnetic field lines tells us that [3,9]

$$\vec{B}_0 = \vec{B}a^2. \tag{6.1}$$

Here $\vec{B}_0$ is the present magnetic field where the scale factor is unity and $\vec{B}$ is the magnetic field when the scale factor was a. Once amplification and diffusion are taken into account, this relationship can be used to calculate the amplitude of magnetic seed fields. Seed magnetic fields produced by Inflation are predicted to be somewhere between $10^{-11}G$ and $10^{-9}G$ on a scale of a few Mpc [3,9,11]. Magnetic seed fields generated by phase transitions are believed to be less than $10^{-23}G$ at galactic scales [3,9]. Some Turbulence theories imply that magnetic fields were not generated until after the first stars were formed therefore requiring no magnetic seed fields [3].

Given how little is understood about primordial magnetic fields and the general lack of agreement among theoretical predictions, it seems clear that the existence of primordial magnetic fields can neither be confirmed or ruled out. According to Dolgov [4] the models involving inflation are the most likely. It seems that the best we can do is set an upper limit on the strength of primordial magnetic fields and utilize this limit in developing models of cosmic structure formation. Observations of the CMB limit the intensity of the magnetic seed fields to $10^{-9}G$ [3,9,11,14].

## 7. The Effect of Gravitational Waves On Cosmic Structure Formation

Work by Duez [7], showed that gravitational waves can induce oscillatory modes in a plasma field. According to Shebalin [24,26], sound waves also lead to magnetosonic and Alfvén waves in the magnetofluid, resulting in an increased velocity of linear perturbations. This could lead to turbulence in the magnetofluid and induce structure formation. Here, structure is defined as cosmic magnetic fields, density and temperature variations and secondary relic gravitational waves. The assumption is that gravitational waves in the early universe (sometime after *t = 1 second*) interacted with the primordial plasma field and resulted in Alfvén and magnetosonic modes [7]. These modes then interacted dynamically, possibly resulting in turbulence and structure formation [24,26].

In order to test this theory the author has constructed a General Relativistic Magneto-hydrodynamic (GRMHD) computer code [6] to model both the plasma field and the background space-time dynamically. The initial space-time was constructed in such a way as to mimic the conditions present around *3* minutes after the big bang. We choose to begin our simulation at *t = 3 minutes* because at that time the primordial plasma field appeared to look like a classical plasma field and the amplitude of the gravitational waves where moderate. At this time we postulate that the temperature of the primordial plasma was around $10^9$ *K*. This plasma was composed mainly of electrons, protons, neutrons, neutrinos and photons. The total mass-energy density proposed at this time is believed to have been $10^4$ *kg/m³*. The specific internal energy, *ε*, is calculated from the initial temperature using *ε = 1.45 a $T^4$ /$ρ_m$*. Here $ρ_m$ is the initial mass density set to *3,367 kg/m³*. The initial pressure was calculated using the gamma-law equation of state with *Γ = 4/3*. This value of gamma corresponds to a radiation-dominated universe. The initial electric field, $E_0$, is zero because the observer is co-moving with the plasma field. The initial spatial velocity of the plasma is zero.

In order to complete the set of initial data, future experiments will also utilize explicit calculation of the viscosity and magnetic field strength for the initial time. These will help determine quantities, such as the dampening scale of the turbulent plasma. This requires further theoretical work, since no universally accepted values are available for this epoch. Based on the available literature [3,4,9,12,15], we find that the initial seed magnetic field should be less than or equal to $10^{-9}$ *G*. This is believed to be a reasonable value of magnetic seed fields for the scale of our simulation generated during inflation. We did not include any viscosity or dissipation in this initial study. Although this magnetic field may seem weak, the fact that it is non-zero helps the gravitational waves to couple with the plasma field and induce oscillatory modes. The purpose of this simulation was to see if the concept proposed here is sound. Future work will start around *t = 1 second* and include scalar perturbations as well as tensor perturbations.

Before running the experiment we thoroughly tested our code. The Duez paper [6] suggested four tests of a GRMHD code, however because of the limited scope of this experiment we felt that only the following two tests were necessary: Gravitational Wave-Induced MHD Waves and Consistency with the Standard Model of Cosmology. We did not include tests of Unmagnetized Relativistic Stars, Relativistic Bondi Flow, or Mikowski Spacetime MHD Tests such as shock tests because the spacetime that we are simulating lacks stars, black holes and supersonic flows. A future test of this structure formation theory may extend into the supersonic region and develop high-density pockets, which could make these tests necessary.

The first test that we performed involved generating Alfvén and magnetosonic modes by gravitational waves and comparing the results against the semi-analytic predictions from the Duez paper [7]. We began by using the same initial conditions as we defined for *t = 3 minutes* above. We then added monochromatic standing waves along the z-axis. The polarizations were varied between both the plus and cross polarizations and the direction of the magnetic field was also applied in a variety of directions. For every variation the test proved successful, the analytic and numerical results proved almost identical. We then removed the gravitational waves to see that the pressure of the virtual fluid did in fact balance the gravitational attraction of the fluid elements. After several thousand iterations, there was no apparent expansion or contraction of the spacetime.

For our second test we replaced the Minkowski space-time with a FRW space-time with parameters consistent with *t = 3 minutes* and tested that it evolved consistently based on our understanding of cosmic evolution during the radiation-dominated era. We set the initial scale factor to *$2.8 \times 10^{-9}$* and the intial Hubble Parameter to *$7.6 \times 10^{16}$ km/s/Mpc*. This test again proved successful. At this point we where prepared to add standing gravitational waves with a spectrum consistent to Grishchuk's predictions with both plus and cross polarizations in all three spatial directions. We then ran four simulations with and without gravitational waves and magnetic fields as shown in table 2.

**Table 2.** FRW Data Runs

|      | Gravitational Waves | Magnetic Fields |
|------|---------------------|-----------------|
| FRW1 | Yes                 | Yes             |
| FRW2 | Yes                 | No              |
| FRW3 | No                  | Yes             |
| FRW4 | No                  | No              |

These simulations showed that while gravitational waves and magnetic fields had little impact on the expansion rate of the space-time, they did have some effect on variations in density, pressure and fluid velocity. In analyzing the data, we looked closely at density variations in each run. Here, density variations are defined as the difference between the density at each point and the mean density of the system. We observed that in both runs involving gravitational waves, the trough to crest density variations seemed to increase with time, reach a peak and then stabilize at some slightly lower constant value instead of increasing without limit. These variations tended to be larger in the run FRW1. This initial growth appears to be caused by the sudden introduction of gravitational waves into the homogeneous plasma field. A further analysis of this data showed that several areas with high and low density formed in the system and remained throughout the simulation. The amplitude of these variations is ~$10^{-10}$ which implies a density fluctuation of about one part in ten trillion given a computational domain of two meters cubed. In today's measurements this corresponds to roughly twice the distance from the earth to the moon. Further work is needed to show how results of this code will match up with observations of the Cosmic Microwave Background. While the full extent of these fluctuations and their impact on structure formation requires further study, these results show that they are much more significant than density fluctuations generated by gravitational waves acting on a hydrodynamic fluid.

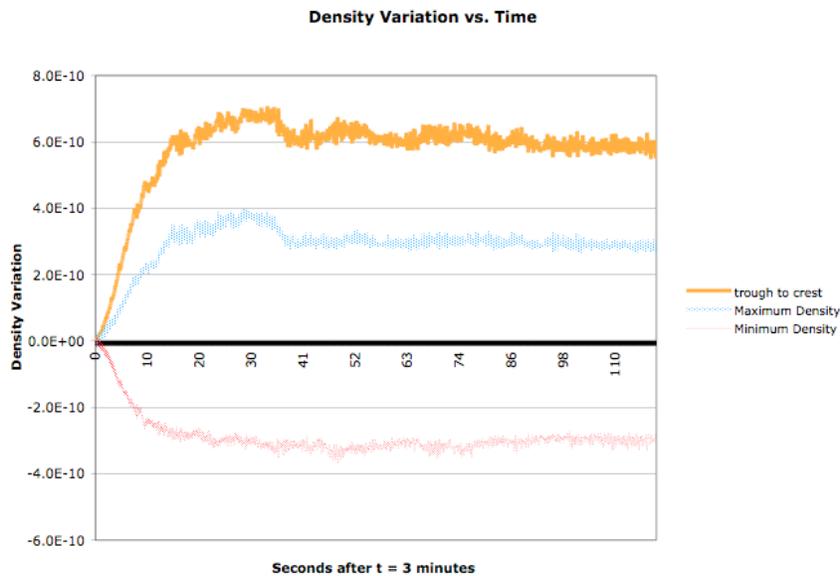

**FIGURE 1**. This figure shows the density variation of the plasma field, observed as a result of an isotropic gravitational radiation field interacting with a homogeneous plasma field using a background FRW metric.

Vortices were also present in the data implying the existence of a turbulent system. This could lead to a dynamo effect that could strengthen the magnetic fields. These preliminary results show that isotropic gravitational waves acting on a homogeneous plasma field, using a FRW background metric, result in regions of increased and decreased density and therefore could lead to structure formation as the spacetime evolves. More work is needed to determine how and if these variations correspond to those observed in our universe. Future work on this topic will include a study of how structure formation depends on scalar perturbations and gravitational waves as well as the viscosity and magnetic field of the initial plasma field.

## 8. Discussion

In this paper, we showed how gravitational waves interact with the primordial plasma field to affect cosmic structure formation. This interaction has been ignored in the past because of the naive assumption that the primordial plasma behaves like a simple hydrodynamic system [13,14,16,23]. When taken in combination, work by Grishchuk [10] and Duez [6,7] demonstrate that it is possible to amplify tensor perturbations into gravitational waves, which are strong enough to induce oscillatory modes in the plasma field. In addition, Grishchuk's work appears to be consistent with the most recent observations of the Cosmic Microwave Background radiation [10]. Work by Alexander shows how gravitational waves may become birefringent further complicating the excitation of modes in the magnetofluid [1]. These modes may become turbulent with or without the addition of birefringence. Shebalin showed how turbulent plasma tends to form coherent structures [24,26]. This implies that the dynamics of the primordial plasma field cannot be ignored in cosmological theory when magnetic fields are present. Preliminary results show that gravitational waves can lead to density perturbations and possibly structure formation unlike a hydrodynamic fluid. Whether or not this additional effect will negate the need for dark matter/energy in structure formation remains to be determined. Other factors such as massive neutrinos may also play a role in structure formation.

## Acknowledgments

The author would like to acknowledge support from the Institute for Space Systems Operations and the University of Houston Clear Lake's Faculty Research and Support Funds. The author would also like to acknowledge the master's degree students who worked on several aspects of this project: Cindi Ballard, David Chow, John Hamilton, Tom Smith, Kevin Depaula, Rafael de la Torre and Marlo Graves. In addition, Dr. Garrison would like to thank Drs. John Shebalin and Samina Masood for many useful conversations and looks forward to future collaboration with them on this project.